# Temporal interference stimulation for deep brain neuromodulation in humans

Pierre Vassiliadis*[1,2], Elena Beanato*[1,2], Maximilian J. Wessel[3], and Friedhelm C. Hummel[1,2,4]

[1] Defitech Chair of Clinical Neuroengineering, Neuro-X Institute (INX), École Polytechnique Fédérale de Lausanne (EPFL), 1202 Geneva, Switzerland.

[2] Defitech Chair of Clinical Neuroengineering, INX, EPFL Valais, Clinique Romande de Réadaptation, 1951 Sion, Switzerland.

[3] Department of Neurology, University Hospital Würzburg, Würzburg, Germany.

[4] Clinical Neuroscience, University of Geneva Medical School, 1202 Geneva, Switzerland.

* these authors contributed equally

**Correspondence:**

Friedhelm C. Hummel, Defitech Chair of Clinical Neuroengineering, Neuro-X Institute (INX), Swiss Federal Institute of Technology (EPFL), Campus Biotech, Geneva, Switzerland and EPFL Valais, Clinique romande de réadaptation (CRR), Sion, Switzerland. Email: friedhelm.hummel@epfl.ch

**Abstract**

For decades, focal non-invasive neuromodulation of deep brain regions has not been possible because of the steep depth-focality trade-off of conventional non-invasive brain stimulation (NIBS) techniques, such as transcranial magnetic stimulation (TMS) or classical transcranial electric stimulation (tES). Deep brain stimulation has therefore largely relied on invasive approaches in clinical populations, requiring surgery. Transcranial Temporal Interference Stimulation (tTIS) has recently emerged as a promising method to overcome this challenge and allows for the first time focal non-invasive electrical deep brain stimulation. The method, which was first validated through computational modeling and rodent work, has now been successfully translated to humans to target deep brain regions such as the hippocampus or striatum. In this *Perspective*, we present current evidence for tTIS-based neuromodulation, underlying mechanisms and discuss future developments of this promising technology. More specifically, we highlight key opportunities and challenges for fundamental neuroscience as well as for the design of new interventions in neuropsychiatric disorders. We also discuss the status of understanding and challenges regarding the basic mechanisms of action of tTIS and possible lines of technological innovation to optimize stimulation, in particular in terms of intensity and focality. Overall, we suggest that following the first proof-of-concepts, an important multidisciplinary research effort is now required to further validate the use of tTIS in multiple applications, understand its underlying principles and optimize the technology in the view of a wider scientific and clinical deployment.

**Introduction**

In the last few decades, non-invasive brain stimulation (NIBS) techniques have attracted significant interest in neuroscience to non-invasively modulate neural activity in specific brain regions. Several technologies including transcranial magnetic stimulation[1] (TMS) and transcranial electrical stimulation[2] (tES), have proved to be efficient tools for fundamental as well as clinical applications[3–7].

In fundamental neuroscience, these techniques allow to alter physiological neural activity of a specific region and to infer its causal role by investigating the subsequent changes in behavior[8,9]. Both TMS and tES have been applied to cortical regions to study their role in a vast range of functions[10,11]. In the last 20 years, NIBS technologies have also shown a promising translational potential[12–14] to support activity when brain regions are hypoactive[15,16], reduce it when it is excessive[16,17] or restore synchronization between brain regions when connectivity is altered[18] (see also [7,16,19,20] for reviews). These insights have allowed approval of some of these protocols for clinical applications by regulatory instances such as the FDA[21]. For instance, TMS is now a validated treatment option for refractory depression[22], addiction[23] or obsessive-compulsive disorders[24] (OCD).

Importantly though, until now, target regions have been largely limited to the cortex because of the limiting depth-focality trade-off characterizing these techniques[25,26]. Stimulating deep brain regions requires higher intensities which ultimately lead to unwanted, off-target modulation of overlying superficial (cortical) regions. For this reason, conventional NIBS approaches do not allow to selectively modulate crucial subcortical areas, such as the hippocampus, the basal ganglia, the thalamus or the cerebellum, which are involved in many cognitive functions and show alterations in multiple neurological and psychiatric disorders[27], including Alzheimer's disease[28,29] (AD), Parkinson's disease[30,31] (PD) or stroke[32,33]). It is important to mention though that some conventional NIBS approaches, such as TMS, can *indirectly* modulate deep brain regions through connectivity, but with arguably limited focality[34–38]. Because of this limitation, the function of subcortical regions has been mainly studied with animal recordings, lesion studies (e.g., after a stroke) or invasive techniques in implanted patients, preventing a detailed exploration and understanding of their healthy functioning. The high prevalence of pathologies involving subcortical regions[39] (e.g., PD, AD, OCD) underscores the need for the development of alternative non-invasive technologies for deep brain stimulation.

A promising method to mitigate the aforementioned challenges may be transcranial temporal interference stimulation (tTIS). The concept, which was introduced for peripheral stimulation more than 30 years ago[40], was recently proposed for brain stimulation and validated through physics and animal experiments[41–43]. tTIS has the potential to overcome the depth-focality trade-off associated with conventional NIBS to allow focal modulation of deep brain structures. It is able to reach deep brain areas, putatively without affecting overlying regions by leveraging the combination of two pairs of electrodes delivering high frequency currents (in the kHz range) with a small frequency difference (Δf) (Figure 1A). The interference between the two currents creates an envelope oscillating at the low-frequency Δf. While the high frequency currents have been proposed to not affect neural activity because they fall outside of the natural range of neuronal operation, the low-frequency beating envelope can influence the regions where its amplitude is high enough to induce neural effects (Figure 1A). Field distribution can be optimized to localize the maximal amplitude modulation in depth, while minimizing it in the overlying tissues. Thus, the overlying tissue will receive mainly unmodulated high frequency stimulation which has been suggested to have minimal neuromodulatory effects (although see "Limitations and current challenges of tTIS" section for in-depth discussion on this point). This optimization can be achieved in two ways, first by the choice of the electrodes placement and second by changing intensity ratio between the two channels. This latter property allows spatial steering of the maximal amplitude without the need of displacing the electrodes[41].

tTIS was first validated in mice by Grossman *et al.* in 2017[41] who showed that tTIS applied at 10 Hz (i.e., by stimulating at 2000 and 2010 Hz) elicited firing of neurons at this frequency, while they remained silent when no interference was induced (i.e., 2000 Hz per channel in antiphase). After stimulation of the hippocampus, c-fos expression (e.g., a marker of neuronal activation) was increased in the target region, but not in the overlying cortex underneath the electrodes (Figure 1B), suggesting selective target engagement. Another study in mice showed that tTIS targeting the superior colliculus, a deep brain region important for eye movement control, evoked eye movements and induced changes of neural activity in the target region[44]. These first rodent studies (see also [45–47]) provided preliminary evidence that tTIS might be able to modulate deep brain regions and induce behavioral changes in a regionally specific manner. Yet, because of the larger head size, translation of tTIS from rodents to human brains is not straightforward[48]. First steps towards larger brains were performed by testing the stimulation on non-human primates[42,43] and suggested that using parameters compatible with human use, tTIS could induce subthreshold modulation of neural activity in deep brain regions (see section

<u>Science: identifying mechanisms of neuromodulation with tTIS</u> for a discussion of mechanisms of tTIS-based neuromodulation).

These first animal findings were corroborated by computational modeling (Figure 1B). Simulations supported the notion that tTIS could induce neural firing in rodents' deep brain regions[49] and that the location of the stimulation could be steered by adapting current ratio between the electrodes[50]. Data from realistic human head models confirmed the reduction of field magnitude when stimulation is applied to larger heads. This prevents direct suprathreshold activation at tolerable intensities in humans, but is compatible with subthreshold modulation[49,51–53]. Individualized human head-brain information extracted from structural MRI also allowed to better investigate and address inter-subject, inter-population variability[54,55], as well as effects of neuronal orientation on fields distribution[56]. These findings underscore the potential value of personalized modeling in optimizing tTIS-based neuromodulation by accounting for individual anatomy to determine electrode placement and current intensity (see section <u>Technology: spatial and functional optimization of tTIS</u>).

Preliminary studies combining tTIS and functional neuro-imaging in humans have provided evidence that tTIS could modulate activity in deep structures in the striatum[57,58] and the hippocampus[59,60] (Figure 1B). These studies targeting deep brain regions highlighted different opportunities offered by the technology, including the application of plasticity-inducing tTIS protocols to strengthen deep brain activity and improve learning[57], the promise of the technique to interfere with specific deep brain rhythms to investigate their causal role for behavior and brain functions[58] and the possibility of steering the temporal interference field in particular deep regions without moving the electrodes[59]. Please see section <u>Fundamental neuroscience: establishing causal links between deep brain activity and behavior</u> for more details on these concepts). Importantly, these applications of tTIS were safe, tolerable and compatible with efficient blinding[61,62]. Multiple other studies also investigated the effect of tTIS on cortical regions in humans[62–66]. This body of work confirmed the feasibility and safety of the technique[62,63] and showed the possibility to modulate different cortical regions, similarly as with conventional transcranial alternating current stimulation (tACS). Overall, this body of work suggests that tTIS is a safe technology that offers the opportunity to selectively and non-invasively modulate deep brain activity with relevant behavioral changes.

Based on these first results, we argue that tTIS represents a new and promising opportunity to (1) widen the understanding on the causal involvement of deep brain regions in specific cognitive functions in humans, (2) investigate the pathophysiology of neuropsychiatric disorders and (3) pave the way to novel treatment strategies to modulate deep brain structures involved in the pathophysiology of, the adaptation to, or the recovery processes from neurological and psychiatric disorders. Notably, it is also important to mention that another technology called transcranial focused ultrasound stimulation (tFUS) is simultaneously emerging for non-invasive deep brain stimulation. As such, tFUS leverages low-intensity ultrasonic waves for focal neuromodulation. This type of neuromodulation, which relies on a radically different concept, has been already covered in several dedicated reviews[67–70] and is beyond the scope of this Perspective. Briefly, we note that current applications of tTIS, while exhibiting a lower spatial resolution than tFUS (in the cm range for tTIS vs. in the mm range in the axis of the transducer for tFUS see also <u>Limitations and current challenges of tTIS</u> section, below), have the advantage of modulating well-studied physiological electrical neuronal mechanisms, exhibit well-defined safety parameters, tolerability and blinding efficiency profiles, and established modeling pipelines, compatible with a large and rapid clinical deployment.

In the present article, we focus on tTIS and highlight promising perspectives and potential challenges for fundamental applications and clinical translation. In the first part of the article, we focus on the applications of the technology from fundamental neuroscience to clinical translation. In the second section, we present some challenges faced by the field, especially regarding putative off-target effects, intensity of stimulation and barriers for clinical deployment. Finally, in the third section, we discuss the possible neural mechanisms of tTIS and present technological optimization options that could improve its effectiveness.

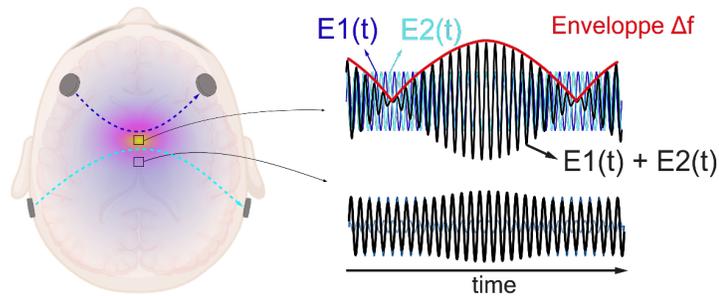

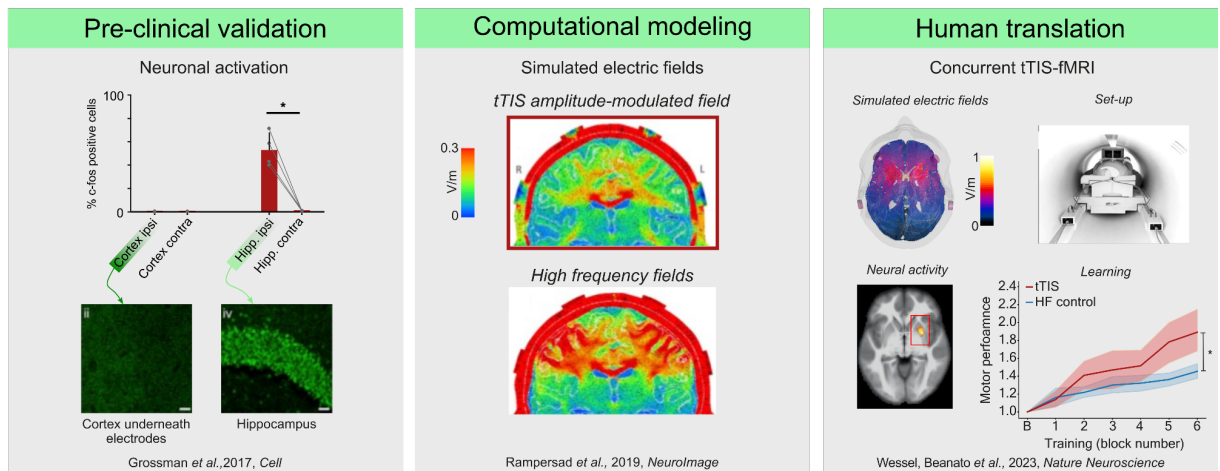

**Figure 1. Concept and translational pathway for tTIS. A. Concept.** On the left, two pairs of electrodes are displayed on a head model, and two currents are applied at frequencies f1 and f1 + Δf. On the right, the interference of the two electric fields within the brain is represented for two different locations with high and low envelope modulation. E1(t) and E2(t) represent the modulation of the fields' magnitude over time. **B. Translational pathway from animals to**

**humans.** On the **left panel**, a figure adapted from a preclinical study shows focal neuronal activation in the stimulated hippocampus and not in the contralateral hippocampus or cortex underneath the electrodes following hippocampal tTIS[41]. On the **middle panel**, a figure adapted from a computational modeling study on a human head model shows electric fields with tTIS as compared to the high-frequency carrier fields[49]. Note that the upper left part shows the temporal interference exposure (amplitude-modulation magnitude) while the lower part shows the high frequency alternating current exposure. On the **right panel**, an adapted figure from a recent human validation study applying tTIS of the striatum during fMRI with concurrent behavior, shows improved motor learning and focal modulation of striatal activity[57]. Please refer to the corresponding articles for more details.

### A) <u>Fundamental and clinical applications of tTIS</u>

1) <u>Fundamental neuroscience: establishing causal links between deep brain activity and behavior</u>

A longstanding paradigm in neuroscience has consisted in the use of lesions to study the causal role of specific brain regions in behavior. In humans, an iconic example of a lesion study is the case of patient H.M., whose bilateral lesions of the hippocampus shed light on the role of this area in declarative memory and shaped many theoretical models of memory function[71,72]. This approach of inferring the causal role of brain regions through careful neuropsychological evaluation of patients with localized lesions has been extremely fruitful throughout the years to understand the neural correlates of various functions including cognitive control[73], executive control[74], social behavior[75,76], language[77], motor control and learning[78–81], motivated behavior[82,83] or time perception[84]. A parallel approach that was introduced by Penfield in the first half of 20th century consisted in the systematic evaluation of a brain region through intracranial electrical stimulation[85]. Modern neuroimaging lesion-symptom and lesion-network mapping approaches now allow to generalize such findings by leveraging the variety of lesion localizations in large populations of patients[86,87] or the effect of invasive deep brain stimulation (DBS) applied on specific implanted brain regions (e.g., in the context of depression[88], or movement disorders[89]). Yet, the conclusions of these studies can be limited by the neural changes associated to the pathology itself or to compensatory mechanisms. Studying patients before and after predictable lesions (e.g., from resective surgery) or anticipated neurodegenerative processes (e.g., due to genetic diseases) may offer a valuable study model to draw conclusions on healthy brain functioning. However, even in these cases, it remains challenging to entirely rule out the influence of the underlying pathology on the observed effects. In general, these paradigms are constrained by the fact that some lesion localizations are more likely than others (e.g., because of the vasculature of the nervous system in the case of vascular lesions) and that DBS is implanted only in very specific regions which have shown therapeutic benefit.

To infer the causal role of specific brain regions in healthy behavior, a popular technique since the 1990's has been to use non-invasive brain stimulation. Exploiting TMS, it became possible to apply a focal and transient perturbation of a brain region and study its impact on neural activity and behavior[90]. By doing so, targeted brain stimulation allowed to satisfy most of the Bradford Hill Criteria of causality, including temporality, specificity, experimental manipulation, counterfactual, dose-response relationship and reversibility[86]. Such perturbations proved very fruitful to study the causal contribution of brain regions in several functions including motor control[91,92], motivated behavior[93,94] or visual perception[95–97]. Notably, while this approach allows for inferring whether a given region causally contributes to a specific behavior, it does not provide information about the subsequent chain of neural events, such as the potential mediating role of other areas within the network. More recently, combining neuromodulation with concurrent neuroimaging, such as electroencephalography[98] (EEG) or fMRI[99], allows to investigate the network effects of a perturbation. Beyond its fundamental interest, this line of research led to the development of repetitive TMS treatments which are now FDA-approved therapies for refractory depression[21], addiction[23] or OCD[24]. However, an important limitation of TMS is that it mainly allows to target superficial (cortical) brain regions because the strength of stimulation decreases with the depth of the target[25,100]. Hence, even if some TMS protocols reach deeper structures[34–37], they induce strong cortical co-activation that can lead to unwanted effects or complicate interpretation of results[101]. A new opportunity provided by tTIS is to non-invasively and focally target deep brain regions to study their causal role in healthy cognition and behavior without the strong cortical co-activation associated with conventional NIBS[58]. Below, we highlight some of the core features of tTIS, including the possibility for spatial steering and timing-selectivity, and discuss how these aspects could enable unprecedented spatio-temporal targeting of deep brain regions.

*Spatial steering*

A first interesting property of tTIS is the ability to steer the location of the maximum electric field envelope modulation without displacing the electrodes. This concept, which is achieved by adjusting the intensity ratio between the two pairs of electrodes (e.g., 2 mA peak-to-baseline per pair or 3mA in one pair and 1 mA in the other), has been initially validated in mice. In 2017, Grossman *et al.* found that it was possible to activate different motor representations in the motor cortex without moving the electrodes, by simply changing the current ratio in the two electrode pairs (see e.g., Movie S1-S3 in [41]). In a more recent human study, Violante *et al.*, (2023) validated the concept with fMRI showing that different parts of the hippocampus could be stimulated by altering the current ratio between the two electrode pairs[59]. This feature of tTIS has some practical implications. First, it allows the comparison of experimental conditions targeting different deep brain (sub)regions without changing electrode positioning, allowing for cross-over designs targeting different brain regions without compromising blinding. This is not possible with conventional tES or TMS approaches as stimulation of control regions requires moving the electrodes or coil, respectively (although see innovative multichannel tES and TMS approaches using electrode and coil arrays, respectively[200, 231]). Second, steering could also allow adjusting the localization of stimulation according to the ongoing activity at a particular training stage (Figure 2B), possibly in a closed-loop fashion. Many behavioral functions involve time-dependent processes such as learning or fatigue, in which neural activity is not static but rather evolves dynamically with the repetition of trials. For instance, previous work suggests that motor learning induces a shift of activity from associative towards sensorimotor striatum, putatively supporting automatization of the learned motor skill[57,102]. By steering tTIS to particular regions of the striatum during learning, it would be possible to ask questions about the causal role of particular subparts of the striatum in different training stages. More generally, this approach could enable stage-specific modulation of deep brain regions according to their involvement in a particular process.

*Timing-selective stimulation*

A second interesting feature of tTIS is temporal selectivity. First, tTIS allows rapid triggering of stimulation without inducing strong skin sensations that are usually experienced with conventional tES technologies when currents are ramped up in less than 5-10s[103]. This is because the high-frequency carriers can be ramped up at the beginning of a task with Δf = 0, and then the interference can be generated by adjusting one of the two high frequency currents to quickly introduce a Δf (as in [57]). This aspect of tTIS makes it appealing for stimulation of very rapid, time-locked processes such as reward processing[104] or for the integration in closed-loop systems in which stimulation needs to be triggered in real-time depending on a particular neural signal. Second, by adjusting the Δf at specific frequencies, it may be possible to modulate specific patterns of oscillatory activity with different temporal dynamics (Figure 2C). As such, a recent paper applying tTIS in non-human primates showed that, consistent with tACS effects[38,105,106], tTIS does not affect firing rates but instead modulates spike timing of individual neurons[42]. More specifically, Vieira *et al.,* (2024) found that when exposed to tTIS, some neurons were entrained but the majority of neurons were desynchronized at the envelope frequency[42]. It is worth mentioning that the propensity to induce entrainment or desynchronization might depend on many factors such as the type of neurons, the initial level of entrainment, the envelope frequency and the intensity of stimulation[42,105] (see section Science: identifying mechanisms of neuromodulation with tTIS for more details). Hence, a promising application of tTIS might be to study the causal role of brain rhythms generated in deep brain structures by modulating coordination of neural firing with a specific frequency of interest (see also section Clinical translation: towards personalized non-invasive deep brain stimulation therapies for clinical applications of this concept). Following this rationale, Vassiliadis et al., (2024) found that application of an envelope frequency of 80Hz in the striatum, a frequency associated to reinforcement learning in the striatum of rodents[107], disrupted reinforcement motor learning compared to the 20Hz control, and that this effect was associated with a suppression of neural activity in the striatum[58]. Consistently, 20 minutes of parieto-occipital alpha tTIS increases event-related alpha desynchronization during a mental rotation task performed after stimulation[108]. Hence, frequency-specific tTIS appears as a promising tool for the causal investigation of specific deep brain rhythms in human behavior. Another promising approach for fundamental neuroscience might be the modulation of plasticity mechanisms in the target region by exploiting classical timing-specific plasticity-inducing (or preventing) patterned protocols (see [109] for a recent study inducing plasticity in the motor cortex of rats using tTIS). Consistently, Wessel, Beanato *et al.,* 2023 showed that striatal tTIS patterned as an intermittent theta burst stimulation (iTBS, i.e., a classical stimulation protocol known to induce long-term potentiation in rodents[110]) increased striatal activity and improved motor learning abilities in healthy adults[57]. It is worth highlighting that tTIS modulated task-induced BOLD changes in the striatum but did not lead to a direct activation of the region at rest. This approach would allow manipulating particular timing-specific plasticity mechanisms in deep brain regions to interrogate their causal role, for instance in various forms of learning processes. Overall, tTIS allows 1) rapid triggering of stimulation compatible with the targeting of transient, time-specific processes; 2) targeting timing-specific deep oscillatory or plasticity mechanisms allowing to infer their causal role in healthy and pathological conditions.

Put together, the steering and timing-selective properties of tTIS enables the targeting of deep neural mechanisms with specific (and possibly evolving) spatio-temporal characteristics to investigate their functional role in behavior. Importantly, such stimulation can be combined with associative neuro-imaging techniques to monitor neural changes associated with the intervention. For instance, three recent studies which have applied tTIS to deep brain regions with concurrent fMRI, showed neural changes in the target region as well as modulation of connectivity with connected brain areas[57–59]. Moreover, some studies have applied cortical tTIS and evaluated the lasting effects of stimulation with electrophysiology (EEG or magnetoencephalography (MEG)) before and after the intervention[64,108]. Efforts are also made to measure electrophysiological activity during tTIS, exploiting a specific set of filters aiming at removing stimulation artifacts[111]. In theory, combination with other neuro-imaging techniques, such as magnetic resonance spectroscopy, or positron emission tomography may also allow a better understanding of the neural effects of tTIS. We suggest that combining tTIS with sophisticated behavioral approaches and multimodal imaging techniques will be a key step forward to better understand stimulation effects and ultimately further establish the technique in human neuroscience.

Overall, tTIS offers a unique opportunity to non-invasively and focally stimulate deep brain regions to determine their causal role in healthy humans, overcoming some of the main limitations of invasive DBS or cortical NIBS. Stimulating deep brain regions with spatially-steered and timing-selective tTIS may generate new insights into the causal spatio-temporal mechanisms of a variety of behavior including learning, motivation, memory formation or executive control.

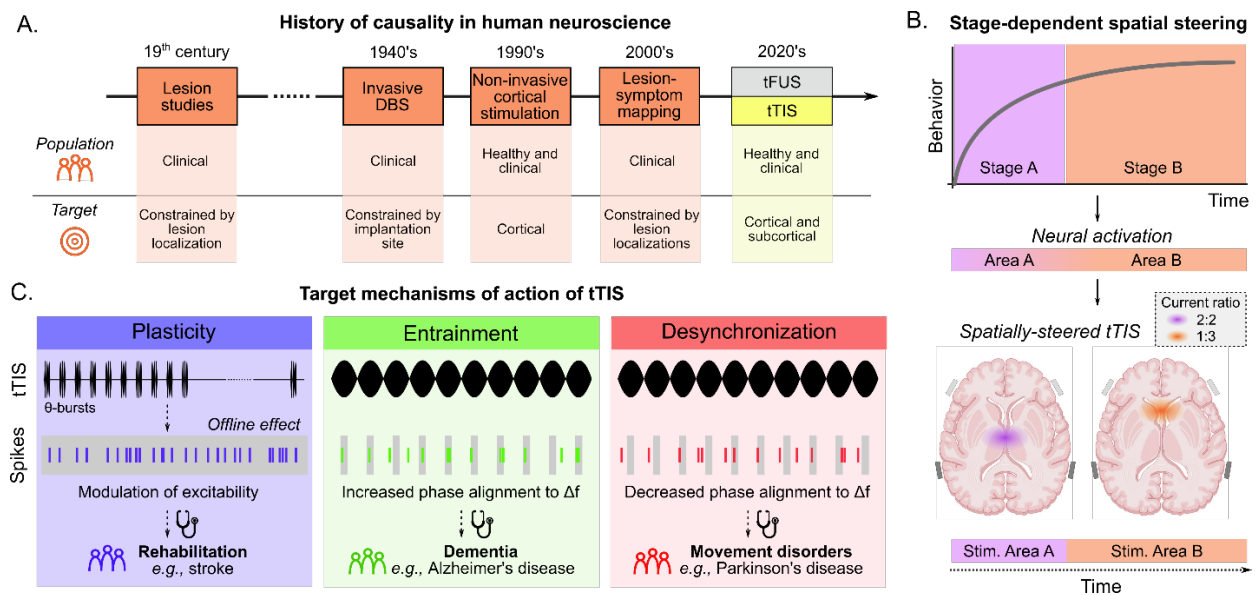

**Figure 2. Opportunities for fundamental and clinical applications. A. A brief (non-exhaustive) history of causality in human neuroscience.** Different experimental approaches and technologies have been used since the 19th century to better understand causal brain-behavior relationships. These methods vary in terms of the populations that can be studied (healthy subjects or patients) and the brain regions that can be investigated (cortical or

subcortical). For the first time, tTIS allows focal, non-invasive, electrical brain stimulation of cortical and deep brain regions in healthy subjects and patients. Note that tFUS is another promising approach for non-invasive deep brain stimulation that relies on ultrasonic neuromodulation with radically different mechanisms (not covered in this perspective). The mentioned dates correspond to the approximative introduction of the technique in humans. **B. Concept of stage-dependent spatial steering.** Consider a dynamical behavioral process such as learning of a new task in which neural activity is not static but rather varies with practice. tTIS allows to target different brain regions depending on their pattern of activity during a specific behavioral stage (e.g., early learning or plateau of performance), without moving the electrodes. This could enable the exploration of new questions regarding the causal dynamical role of specific deep brain regions in certain behaviors. Note that different envelope frequencies could be used to target deep brain regions at their preferred frequency of operation in a stage-dependent way.

**C. Target mechanisms of neuromodulation with tTIS and possible clinical applications.** Three different mechanisms of action of tTIS are proposed. Specific theta-burst patterned stimulation may induce plasticity in the target regions and thereby modulate excitability of the target region (left panel, see [57]). Stimulating at a specific envelope frequency may induce entrainment (middle panel, see [59]) or desynchronization (right panel, see [58]) depending on multiple factors such as the target region, the type of neuron or the initial endogenous pattern of activity at this frequency[42]. We highlight that these three different concepts could lead to different clinical applications, in rehabilitation, dementia or movement disorders.

2) <u>Clinical translation: towards personalized non-invasive deep brain stimulation therapies</u>

The tTIS technology may contribute to translational clinical research across several areas of application. These range from investigations on pathophysiological mechanisms of neuropsychiatric disorders to the development of innovative therapeutic strategies.

*Safety of tTIS*

A crucial step in developing new neurotechnology is ensuring its safety and tolerability. Several studies have explored this issue at various levels, ranging from cellular markers and computational modeling to the assessment of adverse events and biomarkers of pathology. At the cellular level, the results of Grossman et al., (2017) supported the view that tTIS does not induce apoptotic activity, DNA damage or immune response when applied in mice[41]. In humans, Vassiliadis et al., (2024) did not report any adverse events and found that tTIS was well tolerated (mostly tingling sensation perceived as "mild", similar to conventional tES) when targeting deep brain regions (i.e., striatum and hippocampus) in more than 250 sessions[61]. Moreover, Piao et al. (2022) did not find any evidence for epileptic EEG activity nor any increase in neuron-specific enolase (a biomarker of neuronal damage) following cortical tTIS[62]. Previous tACS studies also support the safety of high-frequency electrical stimulation, at least up to 10kHz in humans[112,113]. Finally, guidelines for safe tTIS applications have been recently published based on extensive assessments of electric field simulations and related parameters, namely current density and heating. Based on these metrics, tTIS was suggested to be safe for current intensities up to 3–5 mA and 15–30 mA peak-to-baseline for envelope frequencies in the typical tES and DBS range, respectively[114]. Hence, these first studies suggest that tTIS is safe and well-tolerated supporting its use in larger clinical trials. Still, many unknowns remain regarding other aspects such as the onset time of action, duration and possible reversibility of the effects. Combining tTIS with neural recordings, such as EEG or MEG, is a promising approach to gain further knowledge on these questions. Moreover, we note that more work is needed to establish the safety of tTIS for longer exposures (e.g., applying tTIS over multiple weeks) since most of the current evidence comes from a limited number of sessions (e.g., see [61,62]).

*Blinding efficiency*

A second important aspect for clinical translation is to ensure that tTIS is compatible with efficient blinding. In this regard, the implementation of an appropriate control condition - in most cases a sham (often a quick ramp-up and ramp-down of high frequency currents) or high-frequency control - is pivotal to clinical research and the gold standard for evaluating the efficacy of novel treatment strategies. In a first data set with more than 100 participants and 250 sessions, tTIS of deep brain targets ($striatum\ or\ hippocampus$) showed robust blinding efficiency with respect to both sham and active (high-frequency) control stimulation[61] (i.e., high frequency stimulation without a frequency shift across stimulation channels). This provides an important basis for future translation of tTIS to clinical applications.

*Preliminary clinical data*

Preliminary data on tTIS in patients is now available. Liu et al., (2024) showed that 130 Hz tTIS of the substantia nigra region could reduce tremor scores assessed by the MDS-UPDRS Part III in two patients with Parkinson's disease (PD) and one patient with essential tremor following a double-blind design and testing a 130 Hz tACS control condition[43]. It is noteworthy that the approach addressed an unusual target when compared to invasive DBS. Subsequently, studies were performed on more classical DBS targets. Yang et al., (2024) could show within a randomized, double-blind, sham-controlled study in patients with PD (N = 12) that 20-min 130 Hz tTIS targeting the right globus pallidus internus (GPi) can reduce motor symptoms, specifically bradykinesia and tremor quantified by the MDS-UPDRS Part III[115] (improvement 6.64 points or 14,7 %). In addition, Yang and et al., (2024) reported positive effects of 130 Hz tTIS of the subthalamic nucleus (STN) on motor function quantified with MDS-UPDRS Part III in a separately published uncontrolled and unblinded case series of N = 8 PD patients[116]. A recent study, combining tTIS with STN recordings in 8 PD patients, showed that tTIS targeting the STN could reduce pathological beta activity, similarly as invasive DBS[117]. Notably, although these are important first proof-of-concept studies of tTIS in patients with movement disorders, there are still considerable open questions (see the Limitations and current challenges of tTIS section). Furthermore, in a first pilot cohort including N = 15 patients after traumatic brain injury, a recent study demonstrated the feasibility and blinding efficiency of tTIS targeting deep brain regions[61,118]. Besides these first applications, several studies on tTIS in clinical populations are currently ongoing. Up to date, the database clinicaltrials.gov (retrieval date: 4th of March 2025, search term "Temporal Interference Stimulation") lists N = 19 studies including several clinical conditions, such as bipolar disorder, major depressive disorder, addiction, mild cognitive impairment, AD and gambling disorder. Overall, these first data suggest feasibility of applying tTIS in clinical populations[119]. Efforts are now being made to better characterize possible efficacy in various neuropsychiatric conditions.

tTIS shows promise to target several neuronal mechanisms that are dysfunctional in neuropsychiatric disorders. Based on the currently available stimulation protocols, tTIS could be utilized to (i) to modulate neuroplasticity, (ii) to entrain oscillatory activity or (iii) to disrupt ongoing pathological oscillations[119] (Figure 2C). Hence, tTIS could help to further decipher pathophysiological processes in neuropsychiatric disorders that are characterized by abnormal oscillatory activity, sometimes referred to as "oscillopathies"[120–122], but also to be applied as a therapy to alleviate neurological symptoms by intervening on abnormal brain activity. For this purpose, tTIS is a promising addition to the methods available to date, as it enables non-invasive modulation of deep brain structures and can help investigate causal brain-behavior relationships due to its interventional properties. Thus, tTIS complements other NIBS methods, which will however continue to play an important role due to their own important features, such as ease of use and established protocols for supervised home-based use in the case of transcranial direct current stimulation (tDCS)[123,124] or the possibility of supra-threshold stimulation for TMS-based approaches[1]. In the next section, we will discuss a set of exemplary use cases.

*Use-case 1: tTIS to target dysfunctional plasticity*

A pathological brain state can be characterized by dysfunctional neuronal plasticity across the brain network. Several well-studied model disorders show this feature, such as stroke, dystonia, depression, schizophrenia, addiction, post-traumatic stress or autism spectrum disorder[125–128]. Recently developed patterned tTIS protocols allow to mimic established protocols for plasticity induction in hippocampal slice preparations[129], such as intermittent or continuous TBS paradigms[57,109,130]. The idea here is to use specific patterns of stimulation to bidirectionally modulate synaptic plasticity possibly by promoting cellular mechanisms including long-term potentiation (LTP) or depression (LTD) depending on the specific pattern of stimulation. In addition, the approach may have the potential to remodel abnormal brain network interactions[130].

First, TBS-patterned tTIS can be used to elucidate the pathophysiological role of a particular state of plasticity. One use case could be to study the functional role of activity in key deep brain regions during motor recovery following a stroke. In the past, multimodal approaches using functional neuroimaging and non-invasive brain stimulation have played a crucial role in generating hypotheses about the plasticity of cortical networks and in developing novel stimulation strategies[12,33]. tTIS now opens up the possibility of extending these investigations to important deep brain hubs such as the striatum or thalamus, and may allow the extension of available cortex-centered pathophysiological models.

Second, TBS-patterned tTIS could be used to enhance training effects during motor rehabilitation. Compared to conventional techniques such as tDCS, tTIS has the advantage that it may selectively reach important areas of the motor learning network located deep in the brain, such as the striatum. This is critical because the striatum plays a key role in regulating the transition through different stages of motor learning[102,131]. Here, the steering capabilities of tTIS potentially allow for a precise, training-phase specific application of stimulation (Figure 2B), e.g., by steering the stimulation focus from associative to sensorimotor subregions of the striatum during the course of learning and recovery. This strategy is customizable and holds promise for stabilizing the treatment effects of neurostimulation-assisted motor training across patients.

*Use-case 2: tTIS for entrainment of reduced oscillatory activity*

A pathological brain state can also be characterized by reduced oscillatory activity. One potential use case is reduced gamma band activity in AD[132,133]. It has not yet been fully resolved whether these changes in gamma activity play a causal role or are rather an epiphenomenon. However, optogenetic gamma stimulation (40 Hz) in an AD mouse model was able to reduce the accumulation of amyloid-β protein, which is a key pathological feature of AD[134]. The result was corroborated by a first clinical study showing that chronic 40 Hz light and sound stimulation can reduce brain atrophy, improve functional connectivity patterns and improve performance in a memory recall test in patients with mild AD[135]. Furthermore, it has been shown that 40 Hz tACS of the parietal cortex can improve memory performance and cholinergic neurotransmission in MCI-AD patients[136]. Importantly, reduced gamma rhythms in deep brain regions such as the hippocampus have been described in rodent models of AD[137]. Hence, gamma tTIS of the hippocampus or other deep brain regions involved in AD pathophysiology such as the amygdala[138] appears as a particularly promising line of research. An alternative approach in which the entrainment strategy could be applied is to reinforce beneficial oscillatory patterns during specific task-relevant states. For example, by investigating the functional role of enhanced theta oscillations during the encoding and retrieval of episodic memories, and their subsequent enhancement in therapeutic contexts[59,139]. More generally, the non-invasiveness of tTIS, when compared to DBS, could facilitate the implementation of cross-species translational study designs including healthy subjects and patients with neurodegenerative diseases, potentially creating a "fast track from bench to bedside".

*Use-case 3: tTIS for desynchronization of excessive oscillatory activity*

Pathological brain states can also be caused by exaggerated oscillatory activity. One example for such a case is elevated beta activity in PD[140,141]. Specifically, oscillatory beta activity recorded from the subthalamic nucleus (STN) of PD patients has been linked to the magnitude of akinetic rigid symptoms[142,143] and is reduced by treatment (levodopa or invasive STN-DBS)[144,145]. Interestingly, a recent study on a small cohort (n=8) combining tTIS with STN recordings showed that tTIS targeting the STN could reduce pathological beta activity, similarly as established treatments of PD[117]. Pathological beta activity in PD does not appear to be a unique feature of the STN, rather it is assumed that it propagates throughout the cortico-basal ganglia network[146,147]. Furthermore, first causal brain-behavior relations were demonstrated by non-invasive (tACS) and invasive (DBS) beta stimulation of different brain targets[148,149]. In particular, tTIS extends the previous methodology as it allows to non-invasively examine both cortical and deep nodes of the network under investigation. The approach could allow to perturb and study oscillations linked to deep brain regions in patients without implanted DBS systems, e.g., in early stages of the disease, in patients who are not eligible for invasive DBS or in healthy control subjects. As mentioned above (see section <u>Fundamental neuroscience: establishing causal links between deep brain activity and behavior</u>), a further area of application for tTIS could be to target precisely timing-specific processes (e.g., stimulating at a specific phase of a target oscillation), particularly in the case of brain state-dependent interference protocols. This could be done by triggering tTIS with a rapid adjustment of Δf[57], which eliminates the need for long ramp-up and ramp-down intervals associated with other tES approaches such as tACS or tDCS[150], opening new opportunities for closed-loop non-invasive deep brain stimulation.

### B) Limitations and current challenges of tTIS

Despite the considerable enthusiasm surrounding the development of tTIS since the first proof-of-concept and the significant new fundamental and clinical opportunities mentioned above, the tTIS technology still bears some challenges that we discuss below.

*Lack of suprathreshold stimulation*

It is important to note that like other tES methods (e.g., tACS, tDCS or transcranial random noise stimulation), the intensities used for human applications of tTIS can only induce subthreshold, and not suprathreshold, neuromodulation[52]. This is because tES safety and tolerability is established for intensities of up to a few mA (4 mA peak-to-baseline for conventional tES[151]), which leads to electric field magnitudes < 1 V/m that can only result in subthreshold neuromodulation. Notably, there is preliminary evidence that kHz stimulation could be safe and pain-free for intensities up to 10mA[152]. Yet, even these intensities are unlikely to generate suprathreshold neuromodulation, which requires fields in the range of 10-200 V/m[52,153]. Hence, an inherent limitation of tTIS with respect to invasive (e.g., DBS) or non-invasive alternatives (e.g., tFUS and TMS) is that it cannot induce suprathreshold stimulation in humans. Notably though, selective suprathreshold stimulation of deep brain regions is not possible with TMS and feasibility of suprathreshold stimulation with tFUS has yet to be established in humans[154].

*Possible off-target effects induced by the high frequency carriers*

An important idea underlying tTIS is that the amplitude-modulated field can modulate brain activity with minimal off-target effects. In particular, off-target effects could originate from possible neuromodulation induced by the high frequency carriers that are strongest at the cortical level[49] (see also Figure 1B). Importantly, in 2017, Grossman et al., applied tTIS at suprathreshold intensities on rodents and found that it was possible to selectively induce hippocampal firing without affecting the overlying tissue[41]. Notably though, this conclusion was based on electrophysiology and c-fos-associated neural activation that may not have captured subtle changes of neural activity in superficial layers. Consistently, computational work on models of peripheral axons and cortical neurons suggested that when applied at suprathreshold intensities, tTIS could in theory generate off-target effects through conduction blocks induced by the high frequency carriers[51,52]. Yet, suprathreshold tTIS is not feasible in humans as it would require electric fields of hundreds of mA causing tolerability and safety issues[52].

The question then arises as to whether the high frequency carriers can modulate neural activity when tTIS is applied at subthreshold intensities, as typically recommended in humans (generally ≤ 4 mA peak-to-baseline for conventional tES[145]). Simulations on realistic models of cortical neurons suggest that conduction blocks are unlikely to operate at subthreshold intensities, as they require electric fields 4 to 5 orders of magnitude above electric fields commonly applied in humans[52]. Some experimental findings suggest that low kHz magnetic fields can modulate neuronal activity[155–159]. Yet, even though neuromodulation via magnetic fields is thought to also rely on electric field induction, it is worth noting that effects could differ with respect to tES because of the different characteristics of the generated electric field typically elicited by these different methods (e.g., in terms of focality, directionality, intensity or waveform shape)[159,160]. When it comes to tES, one study has tested the influence of low-kHz frequencies on cortical excitability. On 11 healthy participants, Chaieb et al., 2011 found that 1, 2 and 5 kHz tACS over M1 could increase corticospinal excitability[112]. We note that this result contrasts with combined tTIS-fMRI studies in humans which did not find any increase of BOLD signal in the cortical regions below the electrodes neither with striatal[57] or hippocampal[59] montages. More data are required to reconcile this apparent discrepancy and better characterize the possible effects of high-frequency electric stimulation.

In the meantime, it is crucial to implement appropriate control conditions to accurately interpret tTIS effects. Whenever possible, we advocate for the use of active control conditions, such as a pure high frequency control[44,57,64], to isolate the specific effects of the temporal interference component. When testing more than two conditions is feasible, a robust approach is to combine a sham condition - such as rapid ramp-up and ramp-down of high frequency currents to mimic tTIS-related sensations[61] - with an active control, such as a pure high frequency stimulation[161], tTIS with a control ∆f frequency[58,66], or tTIS targeting a different anatomical region[59]. This approach helps to rule out potential influences of high frequency carrier signals on observed effects and, depending on the experimental design, enables conclusions about the frequency specificity[57,161] and spatial selectivity[59] of tTIS effects.

*Focality of tTIS*

An additional challenge of tTIS is its focality. Compared to standard TMS, tACS, and tDCS, computational work suggests that tTIS could achieve greater focality for deep brain targets[49,162]. Importantly though, experimental human work directly comparing the focality of tTIS and alternative techniques is still lacking. Moreover, depending on the size of the target regions, tTIS cannot achieve millimeter-level precision. While focality could be further optimized through technological advancements, as discussed below, it is important to compare tTIS with alternative methods for reaching deep brain regions, such as deep TMS or tFUS. Deep TMS involves adapting standard TMS coil designs to improve penetration[25]. These coils can modulate deeper brain regions[163]; however, greater penetration is achieved at the expense of focality[163], limiting precise targeting of structures such as the striatum or hippocampus without concomitant cortical activation.

In contrast, tFUS uses ultrasonic waves with high tissue penetration[164], allowing for modulation of deep brain regions with higher spatial resolution than current tTIS protocols, reaching a focality in the millimeter range in the axis of the transducer. While tFUS achieves greater focality than tTIS, its effectiveness depends highly on individual head properties, such as skull thickness and composition[165]. As a result, tFUS is more sensitive to individual variations, with even a 0.1 mm difference in scalp thickness affecting the final target[166]. This sensitivity to small variations in anatomy and transducer placement may challenge the large deployment of tFUS in the clinical context. In general, the choice of using tTIS with respect to other non-invasive alternatives such as tFUS or deep TMS may depend on the specific application, considering various aspects such as the required focality and intensity of stimulation, the availability of imaging and modeling platform to account for individual anatomies and financial aspects. Efforts are now being made to refine its focality with innovative approaches such as the use of multiple channels or co-activation strategies (see section "Technology: spatial and functional optimization of tTIS").

*Future challenges for clinical applications*

Defining effective protocols is a key challenge in translating tTIS into clinical trials and future applications. It should be noted that due to the different points of action and characteristics of the stimulation modalities, such as pulsed, suprathreshold stimulation in invasive DBS and subthreshold, sinusodal stimulation in conventional non-invasive tTIS, a 1:1 transfer of established protocols does not appear to be the most promising approach. For instance, in the PD use-case mentioned above, it could potentially be more effective to disrupt a pathophysiologically significant oscillatory signature (e.g., exaggerated beta oscillations) using a desynchronizing protocol than attempting to imitate 130 Hz invasive DBS.

Finally, the tTIS technology will likely face some of the same challenges and barriers for translation as other novel neurotechnologies, i.e., healthcare and R&D economics, the lack of understanding of mechanisms of action, regulatory barriers, and finally the acceptance of clinicians and patients[167]. This is what made the process of bringing new neurotechnologies to market in the past often inefficient. For example, Schalk et al., (2024) estimate that the go-to-market costs for a non-invasive device are in the range of 10 million US$ over 4 years and for an invasive product in the range of 100 million US$ over 10 years[168]. Furthermore, many neurotechnology-based treatment approaches have a low adoption rate, e.g., less than 15% of eligible PD patients receive

DBS therapy[167]. There is no single solution to address these challenges. Potential approaches are to conduct fully translational mechanistic studies, to coordinate extensive collaboration initiatives, e.g., through open-source databases, or to strengthen industrial-academic-government collaborative activities.

As highlighted in this section, application of tTIS in humans currently presents challenges in terms of intensity, focality and clinical translation. Hence, to unlock the technology's full potential in fundamental neuroscience as well as in clinical neuropsychiatry, a better mechanistic understanding and further technical refinements are needed.

### C) **Development of tTIS: from mechanistic understanding to optimization**

1) Science: identifying mechanisms of neuromodulation with tTIS

Better understanding the mechanisms underlying tTIS neuromodulation is a crucial step to optimize the technology, but also to identify neuronal mechanisms that can be targeted in pathologies. As mentioned above, because of the skin sensations associated with electrical brain stimulation, the intensities used for tTIS (up to a few mA) can only induce subthreshold neuromodulation. For this reason, here we will focus on mechanisms of subthreshold tTIS as applied in humans and will not discuss other putative mechanisms associated to suprathreshold stimulation occurring when applying tTIS at higher intensities as done in animal[41,169] or computational studies[51,52]. Below we review putative common and specific mechanisms of neuromodulation with tTIS compared to conventional tES approaches.

*Common mechanisms of neuromodulation in tTIS and conventional tES*

Similar to other tES techniques, subthreshold tTIS-based neuromodulation is likely to occur through neuronal membrane polarization. tDCS and tACS have been shown to primarily modify the likelihood of endogenous action potentials and their timing[38,106,170,171] which can have profound impact on synaptic efficacy and network synchronization. Recent computational as well as primate work suggests that a similar mechanism is at play during subthreshold tTIS[42,172]. Vieira *et al. (2024),* found that tTIS biased the timing of neuronal firing but did not modulate average firing rates. More specifically they reported that while some neurons were entrained at the envelope frequency, the majority of neurons were rather desynchronized and that this effect depended on the baseline level of entrainment of the neuron[42], as with tACS[105]. Hence, an important mechanism of tTIS could be modulation of entrainment/desynchronization at the envelope frequency that depends on the initial pattern of activity of neurons. Relatedly, the relative phase shift between the endogenous rhythm and the applied stimulation may also influence the effect of tTIS, as for tACS[173]. Other parameters could also influence such effects as the type[174–176] and morphology[177] of the target neurons, their orientation with respect to the electric field[178] and the density of specific neuronal compartments (e.g., soma or axons) in the targeted region[179]. Overall, these elements suggest that like other forms of tES, the sensitivity to tTIS may depend on the initial endogenous pattern of activity in the target region (see [57] for consistent results in humans), but also on structural features of the targeted area. We believe that a particularly promising line of research will be to investigate the relationship between endogenous activity (potentially elicited by a task) and responsiveness to tTIS, using electrophysiological methods such as calcium-imaging or stereo EEG in pre-clinical models[43] and scalp EEG/MEG in humans[64,108,111]. Such data would enable to understand which brain states are more likely to be responsive to tTIS and open the way for functional steering of stimulation effects towards areas that exhibit particular patterns of activity during a task, ultimately improving spatial resolution of tTIS[57] (see section Technology: spatial and functional optimization of tTIS).

*Specific mechanisms of tTIS-based neuromodulation*

Beyond these shared features with other tES approaches, tTIS-based neuromodulation is also thought to rely on specific mechanisms that are due to the particular response to amplitude-modulated high frequency fields. First, modeling as well as experimental work suggests that even though tTIS can generate neuronal effects for clinically-relevant electric fields (~0.4-1V/m), those effects are potentially weaker (at the cortical level) than with conventional tACS[42], especially with increasing carrier frequencies[172,180]. A combination of mechanisms could explain this. First, biological tissues between the electrodes and the brain (e.g., skin, skull) exhibit a frequency-dependent reduction of conductivity[42]. Hence, shunting of current through these tissues is larger for tTIS than for conventional low frequency tES methods, reducing the injected charge on target. Second, in contrast to conventional tACS, tTIS has a pure high frequency content, that is amplitude-modulated at the low-frequency Δf. Hence, neurons need to demodulate the signal to extract the envelope frequency, a process that can occur through frequency mixing[46]. However, recent evidence shows that this demodulation has a cost: response to tTIS is reduced with respect to tACS (even when accounting to current shunting), possibly because neurons are not able to fully demodulate amplitude-modulated waveforms[42]. Importantly, reduced response to tTIS could be partially counterbalanced by higher tolerability, allowing stimulation at higher intensities[181–183]. As such, a recent study demonstrated higher tolerability with increasing frequencies of carriers, with perception thresholds for 2kHz stimulation being 10 times higher than with standard tACS[182]. Pain thresholds were also much higher with kHz stimulation, reaching 3.8mA peak-to-peak at 2kHz versus 0.5mA at classical 10Hz tACS. Hence, an interesting perspective with tTIS is to leverage this higher tolerability to stimulate at higher intensities, not achievable with standard tACS, even when using local anesthetics[182]. In summary, these preliminary data suggest that the kHz component of tTIS requires higher intensities of stimulation to achieve comparable neuronal effects and efficacy as tACS, an aspect that may be addressed thanks to the higher tolerability of kHz electric fields.

As mentioned above, a crucial mechanism underlying subthreshold tTIS is demodulation of amplitude-modulated high frequency electric fields (but see also [169] for putative alternative mechanisms of suprathreshold tTIS in animals). Recent evidence suggests that this process can occur at the single-cell level through ion-channel-mediated mechanisms[46]. More specifically, responsiveness to tTIS is thought to require an ion-channel-mediated rectification process before being low-pass filtered[51]. Such rectification could occur thanks to a mismatch in temporal dynamics between fast inward, depolarizing sodium influx and slower outward repolarizing potassium currents[51]. Notably, in this framework, a neuron would respond to tTIS if the slope of the amplitude-modulated electrical field is sufficient to generate this inward-outward current imbalance which will ultimately bring membrane potential closer to a threshold for depolarization[184]. Consistently, computational modeling shows that responsiveness to tTIS decreases when time-constants of membrane polarization increase, confirming that a crucial determinant of responsiveness to tTIS is indeed fast membrane polarization dynamics in the targeted neurons[180] as well as gate characteristics such as time constants of the specific neurons[185]. In line with a specific role of sodium channel dynamics in tTIS sensitivity, a recent study showed that pharmacologically blocking voltage-gated sodium channels abolishes tTIS-mediated modulation of membrane potential[46]. This effect is not observed for pharmacological blockade of synaptic NMDA, AMPA, and GABA-A receptors, indicating some degree of selectivity in the cellular mechanisms underlying tTIS. Hence, these first mechanistic results suggest that specific ion channel dynamics at neuronal membranes underlie demodulation of tTIS fields, ultimately allowing subthreshold membrane polarization. An important line of future research will be to characterize neurons that are better demodulators and therefore more responsive to tTIS fields. As such, recent in-vitro work applying suprathreshold tTIS on cortical neurons suggests that different cell types exhibit different responsiveness to tTIS (e.g., excitatory pyramidal neurons respond more than inhibitory parvalbumin-expressing neurons), but that this responsiveness strongly depends on network interactions between inhibitory and excitatory neurons[176]. Better understanding in vivo cell-type responsiveness to tTIS, including at subthreshold intensities, could be achieved using optogenetics (possibly combined with fMRI[186]) in rodents or spectroscopy in humans.

Overall, tTIS-based neuromodulation at subthreshold intensities as delivered in humans is suggested to rely on modulation of membrane polarization that primarily affects neuronal spike timing. In contrast to tACS, responsiveness to tTIS requires a demodulation process relying on the temporal dynamics of ion channels at the membrane that may reduce neuromodulatory effects, but which could be counterbalanced by the potentially higher focality and higher intensities that can be tolerably delivered with tTIS. Understanding factors of responsiveness of specific regions and neuron types to particular tTIS fields is a crucial avenue to further optimize the technology and improve its efficacy.

2) <u>Technology: spatial and functional optimization of tTIS</u>

tTIS was shown to reach deep brain targets with higher focality than standard tACS in animals[187] as well as in humans (demonstrated via computational studies[55,162,188]). Even though tTIS allows good targeting of deep regions, there are several prospects for optimization that could further improve focality by reducing off-target electric fields[162]. Furthermore, studies on both cortical models and primate brains found that tTIS effects are weaker with respect to traditional tACS[42,172,189] and that the intensities required to influence neural activity increase for higher carrier frequencies[50,180]. Based on these observations, technological efforts have been made to optimize tTIS by enhancing stimulation focality and neuronal effects in the target region.

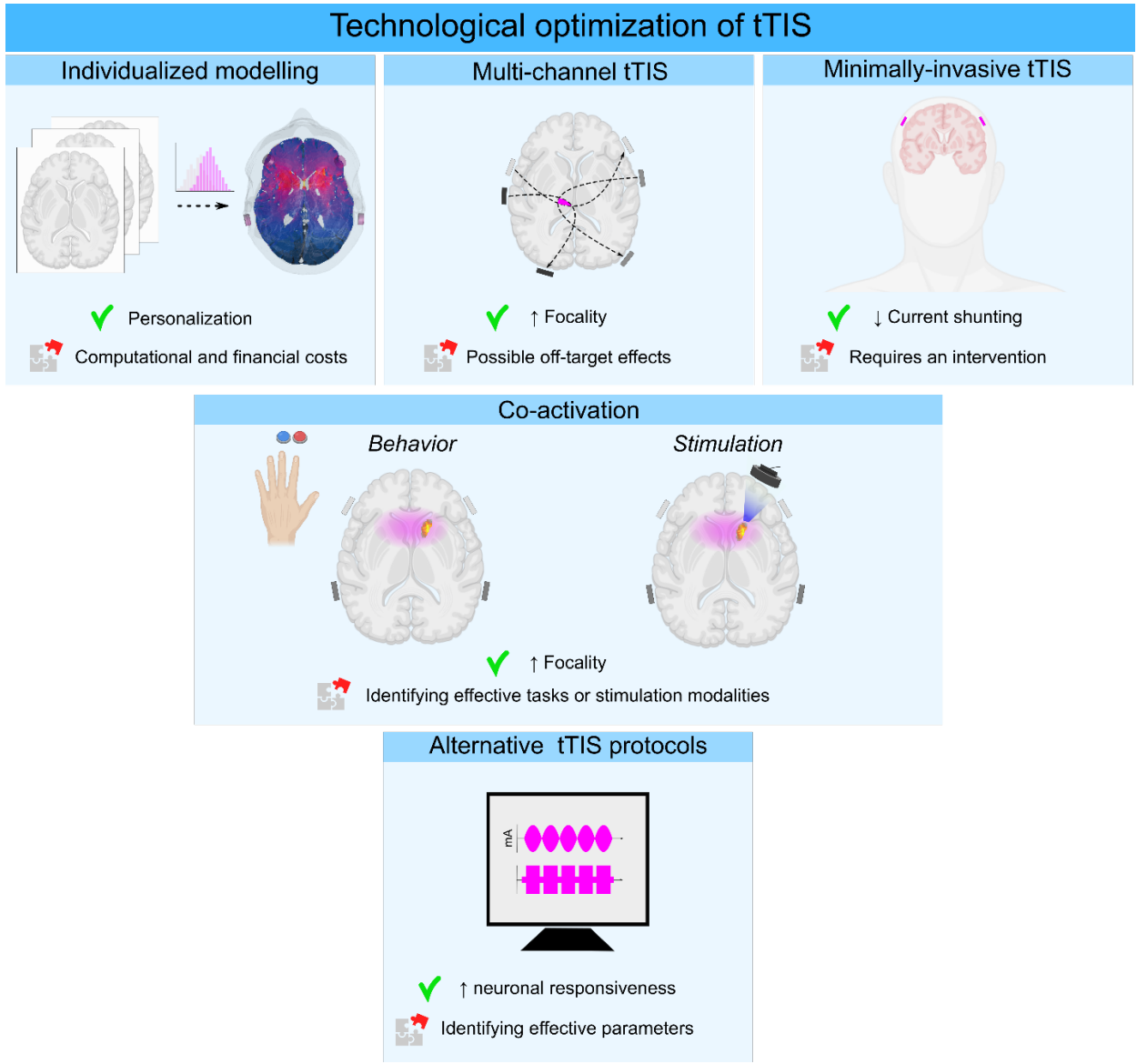

**Figure 3. Technological optimization of tTIS.** Focality of tTIS could be further improved by implementing technological adaptations to currently used tTIS protocols. These optimizations include: (1) personalized modeling on based on individual imaging; (2) increasing the number of electrodes to enhance the amplitude-modulation in a smaller target area; (3) limiting current shunting by using epicranial electrodes; (4) adapting protocol design to pre-activate the target

region via a behavioral task or externally through concurrent brain stimulation (e.g., tFUS or TMS); (5) improving stimulation parameters such as waveforms shape or timing depending on region and task characteristics. Note that these figures are schematic representations and do not reflect a specific result.

*Individualized modeling*

An important factor to consider when it comes to the focality and responsiveness to tTIS is the impact of individual anatomy, morphology and connectivity on the tTIS field distribution[52,54]. Most studies use a predefined set of electrodes optimized for a template head[57–59], which, although capable of achieving significant envelope modulation in the target areas, does not account for variability in individual brain properties. Specific anatomical characteristics could be considered to obtain more precise head modeling by adjusting electrodes size and shape to individual geometries. Population-level studies showed that focality varies between subjects' heads when comparing the electric fields induced by the same montage[54,55]. MR images can provide important information about brain properties such as morphology, level of atrophy or fiber orientation, which can influence field distributions[190] and could hence be used to minimize intersubject variability. Additionally, individual anatomies could also inform electrode designs, which were shown to affect stimulation outcomes in standard tES techniques[191,192]. Current density and field distributions will in fact depend on electrode size[193] and positioning[194,195]. However, modeling personalization has high costs[188,196], both at the logistical and computational level, requiring expensive recordings (MRI) and a high amount of computational time to simulate fields for several combinations of electrodes to find the optimal montage[188]. Therefore, it is essential to investigate to what extent tTIS applications would benefit from the individualization. Some studies found that models based on individual MRI scans demonstrated better target exposure than fields derived from a template[162] and that orientation with respect to the target structure influences tTIS effects[56]. Nonetheless, controversial findings also showed that intersubject variability in the electric fields was lower than the variability of blood oxygenation level, associated with functional brain activity[197], indicating that personalization may not be necessary. Of note, these studies focused on healthy brains. A crucial step for the translation of the technique toward clinical applications will be to explore, not only healthy interindividual differences, but also pathology-related brain changes, such as lesions or atrophy, which are already known to impact on field distributions[198]. Further investigations should address these questions by clarifying the impact of individualized models compared to template-based ones in both healthy and pathological brains, with a particular focus on assessing behavioral outcomes.

*Multi-channel tTIS*

Another interesting concept to improve focality of deep neuromodulation is to apply multi-channel tTIS, exploiting a higher number of pairs of electrodes. Initial investigations were performed with computational modeling and showed that stimulation with more pairs of electrodes decreased off-target fields with respect to standard two–pairs stimulation, with higher focality for increasing electrode pairs[199]. This was confirmed in mice where 6 channels tTIS applied to the primary motor cortex increased focality, reduced sensations and lowered effects of electrode positioning variability[200]. Instead of adding pairs to create fields with comparable properties as the two original channels, a third electrode pair could also be used as a phase-canceling field targeting specific off-target areas to reduce unwanted modulation[201]. Interesting variations of standard multi-channel tTIS were proposed by either using electrode arrays[196,202] or by replacing electrodes pairs with patch pairs able to increase the stimulation precision[203]. However, unintended envelope modulation outside the target was more efficiently reduced when varying additional parameters on top of the electrodes number. Multiple modulation synthesis (MMS) approach was proposed with different channels stimulating at different timings, frequencies and polarities[204]. Of note, for all aforementioned applications, particular attention should be paid to possible unwanted stimulation induced by interfering fields between the additional fields introduced in the head[202], and possible shunting effects due to proximity of the electrodes[205]. For this reason, specific algorithms should be used to find a trade off in the number of electrodes leading to high focality, low computational time, whilst reducing off-target modulations and shunting effects, and maintaining feasibility of the mounting[196,202].

*Minimally-invasive tTIS*

A different approach to increase the field magnitude in the target region would be to use minimally invasive tTIS with electrodes placed subcutaneously to induce epicranial stimulation[206]. This would avoid current attenuation due to skin and muscles, which is one of the main factors leading to shunting effects[48]. This is particularly important for tTIS as there is evidence for a frequency-dependent increase of such shunting[42]. Subcutaneous placement would also increase consistency in electrode locations, allowing repetitive and longer stimulation sessions[207]. Epicranial stimulation has been first tested in animals to apply pulsed electric currents on cortical targets and demonstrated a good safety profile[208], as well as electric fields one order of magnitude larger than tES with standard electrodes on the skin[209]. A first application in humans corroborated safety and efficacy of this approach on epileptic patients with no long-term complications[207,210]. Importantly, electrodes depth and tissue contact are two variables which were shown to be important to reduce current shunting[211]. To increase electric field strength within the brain, the conductive part of the electrode should be placed at the level of the subcutaneous fat layer and avoid contact with the superficial skin layer. Epicranial tTIS represents a promising development of the technique allowing higher intensity, as well as maintaining minimal invasiveness, well suited for clinical applications.

*Co-activation with concomitant behavior or stimulation*

Besides the aforementioned technological development, study design can also be conceived to increase focality of the stimulation and optimize neuronal effects. Due to the subthreshold nature of tTIS in humans, activity modulation in regions at rest is hardly achievable[57,108]. Pre-activation of the target region could enhance effects on this area[212,213], with limited effects in surrounding regions receiving comparable tTIS fields. Behavioral tasks can be used with this purpose, by inducing task-related activity in the target region. Specific paradigms could further boost the focality of the stimulation by selectively activating a subunit of the target region used to optimize electrode placements, without the need of re-running an additional optimization. An exemplary case was presented in a recent study where tTIS applied to the bilateral striatum, increased BOLD activity selectively in one subregion, the putamen, which showed already higher task-related activity during the control condition, and specifically on the contralateral side to the performing hand side[57]. Alternatively, combination with other stimulation techniques such as TMS or tFUS could achieve similar effects by pre-activation of deep structures[214]. TMS has already been used to indirectly target deep structures by stimulating connected cortical regions[215], showing successful influence on network activity[216] linked to associative memory[217], attention[218] or motor control[219]. Modulation of cortical seeds via TMS could induce changes in connected regions of the same network, by either priming deep structures activity or by inducing meta-plasticity that will influence subsequent neural changes[220]. In contrast, direct pre-activation of a deep brain region could also be achieved by tFUS, which is also a promising method to focally reach deep structures[154].

*Alternative stimulation protocols*

Another line of research that has recently started to be explored consists of alternative stimulation protocols to deliver tTIS more effectively. Two fundamentally different protocols already emerged from the current applications: frequency-specific versus TBS-patterned stimulation. The former consists in the delivery of an envelope at a constant frequency which could parallel tACS mechanisms, such as rhythm resonance or entrainment[221,222]. This approach showed divergent neural effects depending on current strength[223], neuron types responses[223] or individual endogenous activity of the target region[105,224]. Even though substantially different mechanisms could underlie tTIS due to the need of extracting envelope frequency (see section above <u>Science: identifying mechanisms of neuromodulation with tTIS</u>), similar considerations could apply, with several parameters of the stimulation protocol to adapt depending on the targeted region and function. An additional challenge in studying tTIS while accounting for endogenous activity is the lack of non-invasive methods to directly record deep brain activity. The second proposed protocol consists of delivering TBS-patterned tTIS to reproduce plasticity protocols (LTP- and LTD- like) initially tested in hippocampal slices[225] and then translated to humans through TMS[226]. While potentially less dependent on the specific frequency, LTP- or LTD- like effects in humans are difficult to investigate and could differ between different brain structures[227,228].

Variations of these two initial propositions have been put forward to optimize stimulation efficacy. A first option to improve temporal control of the interfering fields and create pulse-like envelopes instead of sinusoidal ones could be phase-modulated tTIS[229]. In this version, both electrode pairs deliver the same current frequency, and a transient change of phase is introduced in one pair for a specific duration. To mimic more closely clinical protocols, pulse-width modulated tTIS was also tested in epileptic animal models and showed reduction of epileptogenic biomarkers[187]. In this application, biphasic square pulses are delivered instead of sinusoidal waveforms and showed similar or stronger activity modulation with respect to standard tTIS[230], possibly due to larger energy associated with squared shape. In parallel, multipoint tTIS has been introduced with the goal of increasing the number of stimulation sites in the view of network modulation[231]. Maintaining a 4 electrodes montage, the delivery of two different carrier frequencies per electrode pair and different ratios within each carrier, allowed the creation of two interfering sites, one from the interference of the first carrier frequency and a second from the interference of the second one. This technique would allow multisite stimulation without additional electrodes. Finally, the concept of temporal interference has also been translated to other NIBS techniques such as TMS[232]. First promising results were shown both computationally[233] and in rodents[234]. Nevertheless, further investigations are needed to better understand the mechanisms, effects, and feasibility of TI-TMS. Key discussions - such as focality, montage optimization, and high-frequency effects - are shared between tTIS and TI-TMS. Additionally, feasibility aspects must be addressed, including potential system heating and the coil-dependent decay of the electric field with depth[25]. Finally, possible conduction block effects should be carefully investigated since TMS reaches higher intensities with respect to tES.

*tTIS in combination with recording techniques*

In addition to offering greater focality in deep brain regions compared to traditional NIBS, tTIS also holds potential for concurrent recordings of brain activity. Understanding of NIBS could be elucidated from a network to single neuron resolution perspective by exploiting respectively electroencephalography (EEG), magnetoencephalography (MEG), deep brain recordings (DBS) and in vitro neuron recordings on a chip. The latter has been recently explored in combination with tTIS[235] opening the opportunity to better understand tTIS effects on different types of neurons. In contrast, recording at the macroscale level such as EEG, MEG and DBS are known to be highly affected by major artifacts when combined with traditional tACS, since the frequency of interest during a recording often corresponds to the stimulation frequency. Hence, the signal will be dominated by the injected current frequency, covering natural oscillatory patterns[236]. Because of the high carrier frequencies used in tTIS, above the range of endogenous brain activity, recordings during tTIS are theoretically achievable without artifacts from the stimulation. To note, some practical challenges remain due to the system's nonlinearities which could introduce unwanted artifacts, specifically at the difference in frequency between the two carriers[111]. These artifacts could be mitigated and reduced below noise level via additional hardware exploiting low-pass and high-pass filters, but the design of these components will require testing and adapted protocols to distinguish between brain responses versus stimulation artifacts.

To summarize, based on the promising results of tTIS, significant efforts are being made and need to be made to further optimize the technique, spanning from multichannel solutions, personalization, as well as pre-activation of the target regions to minimally invasive approaches. These advancements will facilitate and open new doors toward clinical translation of the technique.

**Conclusion**

tTIS is a promising method for non-invasive electrical modulation of deep brain regions with good focality, opening new possibilities for both fundamental systems neuroscience and clinical translational applications. We discuss how some of the core features of tTIS including steering without moving the electrodes, rapid triggering of stimulation, the possibility to modulate specific neural processes such as neural oscillations or plasticity mechanisms could be leveraged to answer critical neuroscientific questions and design new interventional strategies for neuropsychiatric disorders. We also present putative mechanisms of action, possible technological developments and outstanding questions for tTIS-based neuromodulation paving the way towards non-invasive precision deep brain stimulation.

**Conflict of interest**

**Acknowledgements**

The research was partially supported by the Akiva Foundation (Crans-Montagna, CH) to FCH, the Bertarelli Foundation (Gstaad, CH) to FCH & MJ, the SNSF Lead Agency (NiBS-iCog 320030L_197899) to FCH, the SNSF Lead Agency (MAPP project 200021E_229225) to FCH, the Defitech Foundation (Morges, CH) to FCH, the Wyss Center for Bio and Neuroengineering (Lighthouse Partnership for AI-guided Neuromodulation) to FCH, the Deutsche Forschungsgemeinschaft (DFG, German Research Foundation, Project-ID 424778381 - TRR 295 A07) to MJW. This work has also been supported by the Social and hUman ceNtered XR (SUN) project that has received funding by the Horizon Europe Research & Innovation Programme under Grant agreement N. 101092612.